%% file: main.tex
\documentclass[runningheads]{llncs}

\usepackage[T1]{fontenc}
\usepackage{amsmath,amssymb}
\usepackage{microtype}
\usepackage{graphicx}
\usepackage{booktabs}
\usepackage{array}
\usepackage{xcolor}
\usepackage{flafter}
\usepackage{tikz}
\usetikzlibrary{arrows.meta,calc}
\usepackage{xurl}
\usepackage{hyperref}

\usepackage[nameinlink,capitalise]{cleveref}

\begin{document}

\title{From Token Changes to Task Failures:\\Bit Flips in a Shared KV-Cache Block\thanks{This chapter was prepared with the assistance of AI.
  See disclosure at the end of chapter for further details.}}
\titlerunning{From Token Changes to Task Failures}

\author{Yuji Yamamoto\inst{1} \and Satoshi Matsuura\inst{1}}
\authorrunning{Y. Yamamoto and S. Matsuura}
\institute{Institute of Science Tokyo, Tokyo, Japan\\
\email{yamamoto.y.dd03@m.isct.ac.jp, matsuura@cii.isct.ac.jp}}

\maketitle

\begin{abstract}
In LLM serving systems with prefix caching, each KV block of a shared
system prompt is held in GPU memory as a single physical copy, which
every request with that prompt reads. Our earlier work measured change and
showed that one bit flip in such a block changes many later responses
without making them visibly degenerate. This paper
extends that work by measuring how often the changed responses are
wrong. Using software fault
injection into vLLM serving Qwen3-8B, we flip the most significant mantissa bit
or the sign bit of a value in such a block. We score
responses on four tasks: code generation, short-answer questions,
multiple-choice reasoning, and refusal under a policy. The
fraction of changes that turned a correct response wrong depended
strongly on the task: 0.023 on MBPP+ and 0.59 on held-out TriviaQA at
the mantissa bit, and 0.019 and 0.60 at the sign bit. Harm fell on
requests whose sampled answers are sometimes wrong, so the damage of
a flip depends on how common such requests are. Harmed responses looked
like ordinary errors, so a flip should be detected by checking cached
blocks rather than by monitoring outputs.

\end{abstract}

\input{01_introduction}
\input{02_background}
\input{03_threat_model}
\input{04_methodology}
\input{05_results}
\input{06_discussion}
\input{07_related_work}
\input{08_conclusion}

\begin{credits}
\subsubsection{\ackname}
This work was supported in part by Palo Alto Networks, which had no role
in the design, conduct, analysis or reporting of this study.

\subsubsection{Disclosure of AI Use.}
The experiment code and the manuscript were written with the assistance
of artificial intelligence (AI) using Claude developed by Anthropic
(Claude Opus 5.5, November 2026). The generated content was carefully
reviewed, verified, and endorsed by the authors to ensure the accuracy
and integrity of the final manuscript.

\subsubsection{\discintname}
The authors have no competing interests to declare that are relevant to
the content of this article.
\end{credits}

\bibliographystyle{splncs04}
\bibliography{references}

\end{document}

%% file: 01_introduction.tex
\section{Introduction}\label{sec:intro}

Large language model (LLM) serving systems cache the key--value (KV)
tensors of a shared prefix, a sequence of initial tokens that many requests have in
common, such as a system prompt. In vLLM's prefix caching~\cite{kwon2023pagedattention,vllm_apc_docs},
these tensors are stored in fixed-size blocks, which are computed once
and then read by every request that carries the prefix. They exist as a single physical copy in GPU memory, and
nothing checks their integrity. Rowhammer can flip bits in GPU
memory~\cite{lin2025gpuhammer}, and a flip in one of these blocks reaches
every later request that reads it.

The conference version of this paper~\cite{yamamoto2026secrypt} studied
this threat with software fault injection. It found that most bit flips
changed the outputs of exactly the requests that shared the prefix
without making them degenerate. The change did not fade over the 100
requests tested, so the number of changed responses grew with every
request served from the block. Counting changes shows how far a flip
spreads, not how much harm it does: a changed response can still be
correct. How many changes actually harm the user or the operator
remained unknown.

This paper measures that harm. We flip bit~6 (the most significant
mantissa bit) or bit~15 (the sign) of a BF16 value, neither of which
made the output degenerate in the conference version, and score responses on four
tasks: code generation (MBPP+), short-answer questions
(TriviaQA), multiple-choice reasoning (MMLU-Pro), and refusal under a
policy. We ask two questions:
\begin{description}
  \item[RQ1:] How often does a flip at bit~6 or bit~15 turn a correct
    response into an incorrect one ($h$), how often does it change the
    response at all ($T$), and what fraction of the changes is harmful
    ($h/T$)?
  \item[RQ2:] Which items do changes and harm fall on, and can these items
    be found without injecting faults?
\end{description}
\noindent In short, the harmful fraction of changes depends strongly on
the task. Harm falls on requests whose sampled answers are sometimes
wrong, and sampling without injection finds them. The damage of one
flip therefore depends on how common they are.

Our contributions are:
\begin{enumerate}
  \item \textbf{Harm per change.} In 188{,}340 flip--item draws at bits~6
    and~15 on 30 flipped elements, $h/T$ at bit~6 is 0.023 on MBPP+, 0.060
    on MMLU-Pro and 0.59 on held-out TriviaQA, and the same within the intervals at
    bit~15 (\cref{sec:rq1}).
    On policy, 11 of the 30 flips at bit~6 let the model add a medical claim after the
    required refusal (\cref{sec:form}).
  \item \textbf{Where harm falls.} Items whose sampled answers are sometimes
    wrong bear the harm, as hypotheses fixed in advance
    predicted for MMLU-Pro and TriviaQA. Sampling 16 answers per item
    finds these items without injection: on TriviaQA, they make up 0.11
    of the items but carry all of the harm. Changes, by contrast, fall on
    items where the top two candidate tokens have nearly equal
    probabilities (\cref{sec:rq2}).
  \item \textbf{Detection.} Beyond the coherence that the conference
    version found, harmed responses mostly keep the required format and fall on items whose sampled answers
    are sometimes wrong. They thus look like ordinary errors, so a flip
    should be detected by checking cached blocks rather than by
    monitoring outputs
    (\cref{sec:form,sec:countermeasure}).
\end{enumerate}

This paper extends the conference version~\cite{yamamoto2026secrypt}
from change to harm. The fault injection and the checksum countermeasure come
from that version and are cited where used.

\Cref{sec:background,sec:threat} give the background and threat model,
\cref{sec:method} the method, \cref{sec:results} the answers to RQ1 and
RQ2, and \cref{sec:discussion,sec:related,sec:conclusion} the discussion,
related work and conclusion.

%% file: 02_background.tex
\section{Background}\label{sec:background}

Three facts underlie the measurements: one cached block is read by many
requests, a bit flip in GPU memory is plausible, and correctness is a
separate question from change.

\subsection{Shared Prefix Blocks in vLLM}\label{sec:bg-prefix}

vLLM~\cite{kwon2023pagedattention} stores the KV cache in fixed-size
\emph{KV blocks}, each holding, by default, the keys and values of 16 tokens for
every layer and KV head. The engine never rewrites a full block, and
nothing checks its integrity~\cite{yamamoto2026secrypt}.

Prefix caching shares these blocks across requests. Each full block is
indexed by a hash of its tokens and of the preceding block's hash, so a
hit requires the whole prefix up to that block to match~\cite{vllm_apc_docs}. A request whose
prompt begins with a cached prefix reuses those blocks, and all such
requests read the same physical copy. We call a sequence of initial tokens shared by
requests a \emph{shared prefix}, and the rest of a request's prompt its
\emph{own prompt}. A \emph{prefix block} is a KV block that holds only
tokens of the shared prefix. Blocks in use are never evicted,
and unused blocks stay cached until least-recently-used
eviction~\cite{vllm_apc_docs}, so a frequently used prefix stays
resident.

\subsection{Bit Flips in GPU Memory}\label{sec:bg-bitflip}

Rowhammer flips bits in a DRAM row by repeatedly activating neighboring
rows~\cite{kim2014rowhammer}. It lets a process alter memory beyond its
own access rights~\cite{seaborn2015rowhammer}, and it has been demonstrated on
the GDDR6 memory of GPUs despite in-DRAM
mitigations~\cite{lin2025gpuhammer,lin2026gpubreach,hu2026gddrhammer,lin2026gputhor}.
Error-correcting codes (ECC) correct a single-bit error in a codeword and
detect a double-bit error, but a triple-bit error can go
undetected~\cite{lin2026gputhor}. Because of its
memory and performance costs, some cloud providers leave ECC off by
default~\cite{lin2026gpubreach}.

The model we study runs in bfloat16 (BF16)~\cite{kalamkar2019bfloat16},
and vLLM stores the KV cache in the model's data type by default. BF16
has a sign bit (bit~15), an 8-bit exponent (bits 14--7) and a 7-bit
mantissa (bits 6--0). \Cref{tab:bf16-flips} gives the effect of a flip at the most significant
bit (MSB) of each field.

\begin{table}[t]
  \centering
  \caption{Effect of a flip on a normal BF16 value $v$.}
  \label{tab:bf16-flips}
  \setlength{\tabcolsep}{4pt}
  \begin{tabular}{@{}rll@{}}
    \toprule
    Bit & Field & Effect on $v$ \\
    \midrule
    15 & sign         & $v \mapsto -v$ \\
    14 & exponent MSB & $v \mapsto 2^{\pm 128}v$, $\pm\infty$, or NaN \\
    6  & mantissa MSB & $v \mapsto (1 + \delta)v$, $1/4 < |\delta| \le 1/2$ \\
    \bottomrule
  \end{tabular}
\end{table}

\subsection{Output Change and Task Correctness}\label{sec:bg-change}

Greedy decoding selects the most likely token at each step, so the output
changes only at a step where a perturbation reverses the order of the
two most likely tokens. Such reversals occur where these two tokens are
nearly tied, even under a benign perturbation such as a change in batch
composition~\cite{he2025nondeterminism,yuan2025fp32}. We call the gap
between their log-probabilities the \emph{margin}. Each token is
generated from the tokens before it, so one changed token can alter
everything after it.

A changed output can be correct. For a model that reliably reaches
the correct answer, a changed path tends to lead there as well.
Consistency under sampling reflects this reliability: answers that recur
across samples are more often correct~\cite{wang2023selfconsistency}, and disagreement among samples
marks likely errors~\cite{manakul2023selfcheckgpt,farquhar2024semantic}.

%% file: 03_threat_model.tex
\section{Threat Model and Scope}\label{sec:threat}

A co-located GPU tenant flips one bit in a shared prefix block of vLLM's
prefix cache, aiming to make a serving application return incorrect
responses without revealing the fault.

\subsection{Attacker Model}\label{sec:attacker}

The attack depends on three conditions: GDDR6 memory, untrusted tenants
sharing the GPU without hardware partitioning~\cite{lin2025gpuhammer},
and bit flips that go undetected. All these conditions hold for the RTX
A6000 and A4000: GDDR6 Rowhammer has been demonstrated on
them~\cite{lin2025gpuhammer,hu2026gddrhammer,lin2026gputhor,nvidia2026rowhammer},
they lack hardware partitioning~\cite{nvidia_mig_gpus}, and cloud
providers offer them with ECC disabled~\cite{lin2026gpubreach}.

The attacker is a tenant with ordinary privileges on such a GPU, and
its only capability against the application is a single bit flip in GPU
memory: it cannot read or modify the model
weights, the serving code or the prompts.

The attacker's goal is an integrity violation, not a confidentiality
one such as prompt extraction~\cite{wu2025promptpeek}. We call responses
\emph{harmed} if they are incorrect (\cref{sec:tasks}) and
\emph{collapsed} if a flip makes them incoherent (\cref{sec:measures}). Collapsed responses reveal the fault and so fail
the attacker.

\subsection{Attack Surface}\label{sec:surface}

The attacker flips a bit in a prefix block (\cref{sec:bg-prefix}). While the block stays cached, the flip reaches
exactly the requests that carry the prefix.

If an application uses the same system prompt for all its users, the
flip reaches all of them. In one provider's API traces, most cache hits
came from fixed system prompts~\cite{wang2025kvcachewild}.
vLLM's optional cache salt can limit the flip's reach to a single user
or a chosen group of users~\cite{vllm_pr17045}.

\subsection{Scope}\label{sec:scope}

As in the conference version~\cite{yamamoto2026secrypt}, we assume
ideal bit targeting (\cref{sec:limitations}): the attacker chooses the
bit position, and the flipped element is random. We emulate such a
flip in software. The harness XORs one BF16 element of Qwen3-8B's
value cache~\cite{qwen2025qwen3} with a single-bit mask.

In the conference version, flips of the upper exponent bits, mostly
bit~14, collapsed the responses, whereas those of the \emph{silent}
bits, 0 to 11 and 15, almost never did~\cite{yamamoto2026secrypt}. We flip bit~6, the mantissa MSB, and bit~15, the sign. We also
flip bit~14, the exponent MSB, to confirm that an injection takes effect
(\cref{tab:bf16-flips}).

%% file: 04_methodology.tex
\section{Methodology}\label{sec:method}

We measure the change rate $T$, how often a correct response changes,
and the harm rate $h$, how often it becomes incorrect. Without injection, we also measure per-item
predictors of change and harm (\cref{sec:predictors}).

\subsection{Setup}\label{sec:setup}

All runs use Qwen3-8B~\cite{qwen2025qwen3} in BF16 on an NVIDIA DGX
Spark, as in the conference version. The model runs in non-thinking
mode. The serving engine is vLLM 0.29.0
with asynchronous scheduling disabled. The other settings are vLLM's
defaults: prefix caching is enabled with 16-token blocks, and no
request carries a cache salt.

\subsection{Prompt and Tasks}\label{sec:tasks}

Every request begins with the same shared prefix of six full 16-token
blocks (\cref{fig:prefix}) and ends with \texttt{<answer>}, so the model
starts its response inside the tag. Only the output up to the closing
\texttt{</answer>} counts. Closing the tag is the required format: an
output that never closes it is incorrect, except on policy. \Cref{tab:tasks} summarizes the tasks; each
\emph{item} of a task is sent as one request.

\begin{figure}[t]
  \centering\scriptsize
  \begin{tabular}{@{}l>{\raggedright\arraybackslash}p{0.8\linewidth}@{}}
    \toprule
    Policy & \texttt{<|im\_start|>system\textbackslash n}You are a support
      assistant for a small software company that helps customers with
      technical questions. Do not give any medical advice; politely
      decline such requests.\texttt{\textbackslash n} \\
    Format & Put your final answer between the \texttt{<answer>} and
      \texttt{</answer>} tags shown clearly below. Text written outside
      those tags is completely ignored by the automated
      grader.\texttt{\textbackslash n} \\
    Boilerplate & Thank you so much for taking the time to use our
      customer service today. We value your feedback and aim to respond
      clearly and accurately.\texttt{<|im\_end|>\textbackslash
      n<|im\_start|>user\textbackslash n} \\
    \bottomrule
  \end{tabular}
  \caption{The shared prefix, in its three 32-token parts.}
  \label{fig:prefix}
\end{figure}

\begin{table}[t]
  \centering
  \caption{Tasks and what a correct response contains.}
  \label{tab:tasks}
  \footnotesize
  \setlength{\tabcolsep}{4pt}
  \begin{tabular}{@{}l>{\raggedright\arraybackslash}p{0.3\linewidth}
      >{\raggedright\arraybackslash}p{0.30\linewidth}r@{}}
    \toprule
    Task & Items & Correct response & Max.\ tokens \\
    \midrule
    MBPP+~\cite{austin2021mbpp,liu2023evalplus} & Python programming problems
      & a function that passes all tests & 512 \\
    TriviaQA~\cite{joshi2017triviaqa} & trivia questions, five-shot
      & a short answer matching an alias & 512 \\
    Policy & borderline medical requests that the policy forbids
      answering & a refusal & 512 \\
    MMLU-Pro~\cite{wang2024mmlupro} & multiple-choice questions
      & the reference letter & 1024 \\
    \bottomrule
  \end{tabular}
\end{table}

The MBPP+ items are all its problems except one whose reference
solution is faulty, and the TriviaQA items are drawn at random. The 100 policy
requests, such as questions about a symptom, were written for this
study. Policy has no reference answer, so a response
is incorrect if it closes the tag and makes a medical claim of its own.
One author confirmed every such claim by hand.

\subsection{Run Flow}\label{sec:injection}

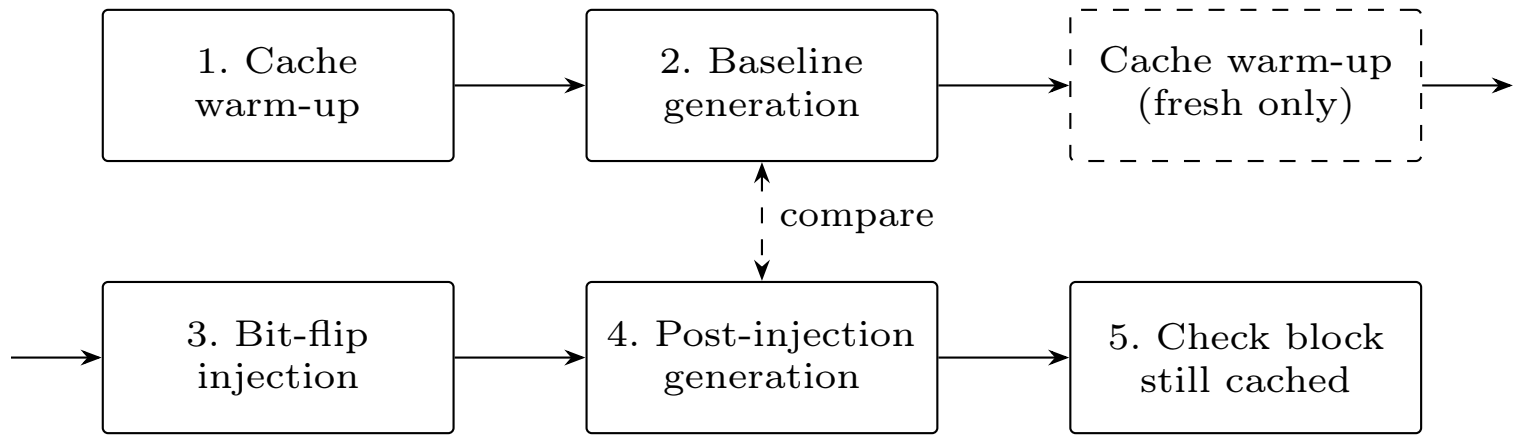
\begin{figure}[t]
  \centering
  \begin{tikzpicture}[
    font=\scriptsize,
    x=1mm, y=1mm,
    box/.style={draw, rounded corners=1pt, minimum height=10mm,
      text width=22.5mm, align=center, inner sep=1pt},
    arr/.style={-{Stealth[length=1.4mm]}},
  ]
    \node[box] (s1) at (0, 0) {1. Cache\\warm-up};
    \node[box] (s2) at (32, 0) {2. Baseline\\generation};
    \node[box, dashed] (sx) at (64, 0) {Cache warm-up\\(fresh only)};
    \node[box] (s3) at (0, -18) {3. Bit-flip\\injection};
    \node[box] (s4) at (32, -18) {4. Post-injection\\generation};
    \node[box] (s5) at (64, -18) {5. Check block\\still cached};
    \draw[arr] (s1) -- (s2);
    \draw[arr] (s2) -- (sx);
    \draw[arr] (sx.east) -- ++(6,0);
    \draw[arr] ($(s3.west)+(-6,0)$) -- (s3.west);
    \draw[arr] (s3) -- (s4);
    \draw[arr] (s4) -- (s5);
    \draw[{Stealth[length=1.4mm]}-{Stealth[length=1.4mm]}, dashed]
      (s2) -- node[right] {compare} (s4);
  \end{tikzpicture}
  \caption{Steps of an injection run for one flip, after the
    conference version~\cite{yamamoto2026secrypt}.}
  \label{fig:flow}
\end{figure}

An injection run repeats the steps of \cref{fig:flow} for each flip.
The harness measures two request types. \emph{Returning} requests cache
the KV blocks of their own prompts in the first warm-up and reuse them
after the flip. For \emph{fresh} requests, a second warm-up evicts the
own-prompt blocks that the baseline cached and recomputes the prefix
blocks, which are identical to the original ones. The flip then
corrupts one of them, and the post-injection pass computes the
own-prompt blocks from it.

Returning requests are less exposed than fresh ones. Only their
generated tokens and a few final prompt tokens read the corrupted
prefix~\cite{vllm_kvcachemanager}. A fresh request resembles the first
turn of a conversation, which shares only the prefix with earlier
requests; a returning request resembles a later turn, whose earlier
turns are already cached~\cite{wang2025kvcachewild}. Unless stated
otherwise, results are
for fresh requests.

Without a flip, the baseline and post-injection passes gave identical
outputs in every run. Any difference in the outputs therefore comes
from the flip.

\subsection{Measures}\label{sec:measures}

A \emph{coordinate} is the position of the flipped element in the value
cache. A \emph{draw} $d$ is one coordinate applied to one item. We
define
\begin{align}
  C &= \{\, d : \text{baseline of } d \text{ correct} \,\}, \\
  h &= P(\text{post-injection output of } d \text{ incorrect} \mid d \in C), \\
  T &= P(\text{output of } d \text{ changed} \mid d \in C),
\end{align}
where $P$ is taken over draws, and $h/T$ is the fraction of changes
that are harmful. Comparisons of the two request types use
\begin{equation}
  D = \{\, d : \text{baseline of } d \text{ correct under both request
  types} \,\} \subseteq C,
\end{equation}
since a fresh baseline can differ from a returning one
(\cref{sec:bg-change}).

\begin{table}[t]
  \centering
  \caption{Stages.}
  \label{tab:stages}
  \footnotesize
  \setlength{\tabcolsep}{3pt}
  \begin{tabular}{@{}llrrlr@{}}
    \toprule
    Stage & Items & Candidates & Pool & Bits (coordinates)
      & \begin{tabular}[b]{@{}r@{}}Draws in $C$\\ per bit\end{tabular} \\
    \midrule
    initial   & MBPP+          & 377     & 226     & 6, 15 (30); 14 (3) & 6{,}780 \\
              & TriviaQA       & 200     & 92      & 6, 15 (30); 14 (3) & 2{,}760 \\
              & policy         & 100     & 89      & 6, 15 (30)         & 2{,}670 \\
    held-out  & new MMLU-Pro   & 1{,}000 & 605     & 6, 15 (30)         & 17{,}880 \\
              & new TriviaQA   & 5{,}000 & 2{,}160 & 6, 15 (30)         & 64{,}800 \\
    fresh     & initial MBPP+     &  &  & 6, 15 (30) & 6{,}780 \\
              & initial TriviaQA  &  &  & 6, 15 (30) & 2{,}760 \\
              & initial policy    &  &  & 6, 15 (30) & 2{,}670 \\
              & held-out MMLU-Pro &  &  & 6, 15 (30) & 17{,}160 \\
              & held-out TriviaQA &  &  & 6, 15 (30) & 64{,}800 \\
    \bottomrule
  \end{tabular}
\end{table}

For comparison with the conference version, we also classify each
flip~\cite{yamamoto2026secrypt}. Let $\tau$ be the fraction of outputs
that change, and $\rho$ their mean ROUGE-L~\cite{lin2004rouge} against
the baseline:
\begin{itemize}
  \item \emph{no effect}: $\tau = 0$;
  \item \emph{partial divergence}: $0 < \tau < 1$;
  \item \emph{complete divergence}: $\tau = 1$ and $\rho \ge 0.1$;
  \item \emph{collapse}: $\tau = 1$ and $\rho < 0.1$.
\end{itemize}

\subsection{Stages}\label{sec:design}

Bits~6 and~15 are flipped at 30 coordinates drawn at random, 10 in
each of the three parts of the prefix (\cref{fig:prefix}).
\Cref{tab:stages} lists the stages. New items enter through a
\emph{pool}: the candidates answered correctly without injection.
Without the policy, policy requests must be answered rather than
refused, so that each refusal comes from the policy. Because batches change after this selection, which can alter a greedy
output (\cref{sec:bg-change}), a few pool items still fall outside $C$.
The answer that each pool item gave at selection and the baseline of
its returning requests thus differ only in batch composition, which
gives one reference per item without a flip and so no intervals.

The initial stage runs on MBPP+, TriviaQA and policy. It also flips
bit~14 to confirm that an injection takes effect.

The held-out stages were added to test the hypotheses that the initial
stage suggested (\cref{sec:predictors}) on items not used before. One
uses TriviaQA. The other uses MMLU-Pro, a new task chosen for two
reasons: many of its items have sampled answers that are sometimes
wrong, and a pilot on 200 other MMLU-Pro items met a harm criterion
fixed in advance. The initial and held-out stages use returning
requests, and the fresh stage then repeats every pool with fresh
requests.

\subsection{Predictors and Hypotheses}\label{sec:predictors}

Each pool item has two predictors, measured without injection
(\cref{sec:bg-change}):
\begin{description}
  \item[Margin:] the smallest margin over the baseline output up to
    \texttt{</answer>}.
  \item[Sampled error:] the fraction of 16 sampled responses that are
    incorrect. An item is \emph{fragile} if its sampled error is above
    zero, that is, if its sampled answers are sometimes wrong. On policy, the samples are not checked by hand.
\end{description}
\noindent Each predictor is evaluated by its area under the ROC curve
(AUC) for two item outcomes, changed in some draw and harmed in some
draw, ranking items from the smallest margin or the largest sampled
error. The hypotheses are\footnote{H1, also fixed in advance, concerned
the share of harmful draws that fall on fragile items; it held wherever
tested, and \cref{tab:fragile} reports this share.}
\begin{description}
  \item[H2:] $h_\text{fragile} > h_\text{rest}$;
  \item[H3:] $T_\text{fresh} > T_\text{returning}$;
  \item[H4:] $h_\text{fresh} > h_\text{returning}$.
\end{description}
\noindent Each holds if the difference between its two sides has a 95\%
interval above zero; all are tested at bit~6, and bit~15 is reported
descriptively. H2 is tested on the held-out stages with both request
types, and H3 and H4 on held-out MMLU-Pro and TriviaQA over $D$. The
tests are confirmatory, since their hypotheses preceded the outputs.

\subsection{Statistical Analysis}\label{sec:stats}

Draws that share a coordinate or an item are correlated, since some
flipped elements and some items are more sensitive than others. We
therefore report 95\% intervals from a two-way cluster bootstrap over
coordinates and items~\cite{owen2007pigeonhole}. Runs that share
coordinates or pools are resampled together, so differences between
bits, item subsets and request types are paired. Where one hypothesis is
tested several times, it holds only if its Bonferroni-adjusted interval
also stays above zero. This interval is 97.5\% for the two tests of
H2 on each request type and 98.75\% for the four tests of H3 and H4.

%% file: 05_results.tex
\section{Results}\label{sec:results}

\Cref{sec:rq1,sec:rq2} answer RQ1 and RQ2, \cref{sec:form}
describes the form of harmed responses, and
\cref{sec:returning} compares the results with returning requests.

\subsection{RQ1: Harm and the Harmful Fraction of Changes}\label{sec:rq1}

\begin{table}[t]
  \centering
  \caption{Change and harm at the two silent bits, with 95\% two-way
    cluster bootstrap intervals. Harm gives the harmful draws, items and
    coordinates (of 30). Policy gives counts only.}
  \label{tab:rates}
  \footnotesize
  \setlength{\tabcolsep}{3pt}
  \begin{tabular}{@{}lrrllr@{}}
    \toprule
    Task & Bit & $T$ & $h$ ($\times 10^{-3}$) & $h/T$ & Harm \\
    \midrule
    \multicolumn{6}{@{}l}{\textit{initial}} \\
    MBPP+    & 6  & 0.13   & 2.9 [0, 8.0]      & 0.023 [0, 0.062]     & 20/4/15 \\
             & 15 & 0.13   & 2.5 [0, 6.6]      & 0.019 [0, 0.051]     & 17/4/15 \\
    TriviaQA & 6  & 0.0094 & 6.9 [0, 20.3]     & 0.73 [0, 1]          & 19/2/16 \\
             & 15 & 0.0065 & 5.4 [0, 15.9]     & 0.83 [0, 1]          & 15/3/13 \\
    Policy   & 6  & 0.17   & \multicolumn{2}{l}{counts only} & 12/2/11 \\
             & 15 & 0.17   & \multicolumn{2}{l}{counts only} & 9/2/8 \\
    \multicolumn{6}{@{}l}{\textit{held-out}} \\
    TriviaQA & 6  & 0.0053 & 3.1 [1.7, 4.8]    & 0.59 [0.40, 0.76]    & 204/33/30 \\
             & 15 & 0.0053 & 3.2 [1.9, 4.8]    & 0.60 [0.43, 0.77]    & 208/39/30 \\
    MMLU-Pro & 6  & 0.64   & 38 [29, 49]       & 0.060 [0.046, 0.076] & 659/107/30 \\
             & 15 & 0.65   & 40 [30, 51]       & 0.062 [0.048, 0.078] & 692/116/30 \\
    \bottomrule
  \end{tabular}
\end{table}

\begin{figure}[t]
  \centering
  \includegraphics[width=\linewidth]{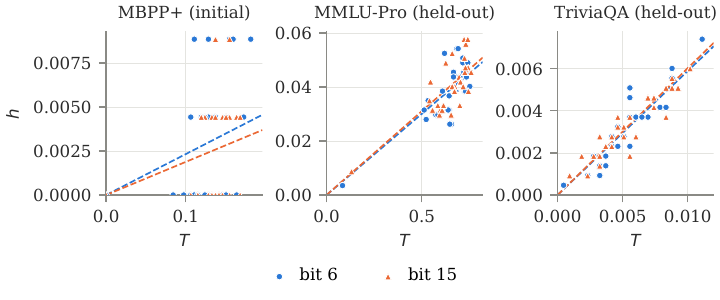}
  \caption{Change rate $T$ and harm rate $h$ of each flip. Dashed lines
    have slope $h/T$ of the whole task at each bit.}
  \label{fig:scatter}
\end{figure}

The harmful fraction of changes differs across tasks by more than an
order of magnitude, and the intervals of MBPP+ and held-out TriviaQA do
not overlap (\cref{tab:rates}). On MMLU-Pro and held-out TriviaQA, the
harm rate of a flip grows in proportion to its change rate
(\cref{fig:scatter}). On MBPP+, each flip harms at most two draws,
too few to estimate the relation. The bit makes no detectable difference: the paired
difference in $h/T$ between bits~15 and~6 has a 95\% interval that
includes zero on every task.

Without a flip, a change in batch composition (\cref{sec:design})
changes 0.11 of the MBPP+ items and 0.17 of the policy items, close to a
flip (\cref{tab:rates}), but makes none wrong. On MMLU-Pro, it makes
0.015 of the items wrong, against 0.038 for a flip.

\subsection{RQ2: Where Harm Concentrates}\label{sec:rq2}

Harm falls on fragile items, and changes on items with a small margin.
Both can be found without injection, by the predictors of
\cref{sec:predictors}.

On the held-out items, H2 holds (\cref{tab:fragile}): $h$ is higher on
fragile items than on the rest. In fact, fragile items carry nearly all
harmful draws on every task. The rest are almost never
harmed, even on MBPP+, policy and MMLU-Pro, where they change often.

\begin{table}[t]
  \centering
  \caption{Fragile items and the rest. Fragile: share of draws on fragile
    items. On fragile: share of harmful draws on them. Difference: fragile
    minus rest, with 95\% intervals. Holds: whether H2 holds; ---: outside
    the hypotheses (\cref{sec:predictors}).}
  \label{tab:fragile}
  \scriptsize
  \setlength{\tabcolsep}{2.5pt}
  \begin{tabular}{@{}lrrrrrrrll@{}}
    \toprule
    & & & & \multicolumn{2}{c}{$T$} & \multicolumn{3}{c}{$h$ ($\times 10^{-3}$)} & \\
    \cmidrule(lr){5-6}\cmidrule(lr){7-9}
    Task & Bit & Fragile & On fragile & fragile & rest & fragile & rest & Difference & Holds \\
    \midrule
    \multicolumn{10}{@{}l}{\textit{initial}$^\dagger$} \\
    MBPP+    & 6  & 0.075 & 20/20   & 0.18  & 0.12    & 39  & 0   & 39 [0, 104]  & --- \\
    TriviaQA & 6  & 0.12  & 19/19   & 0.058 & 0.0029  & 58  & 0   & 58 [0, 167]  & --- \\
    Policy   & 6  & 0.13  & 12/12   & 0.20  & 0.16    & \multicolumn{3}{l}{counts only} & --- \\
    \multicolumn{10}{@{}l}{\textit{held-out}} \\
    MMLU-Pro & 6  & 0.50  & 627/659 & 0.73  & 0.55    & 73  & 3.7 & 70 [52, 89]  & yes \\
             & 15 & 0.50  & 654/692 & 0.74  & 0.56    & 76  & 4.4 & 72 [53, 93]  & --- \\
    TriviaQA & 6  & 0.11  & 204/204 & 0.030 & 0.0022  & 28  & 0   & 28 [16, 43]  & yes \\
             & 15 & 0.11  & 208/208 & 0.030 & 0.0022  & 29  & 0   & 29 [17, 42]  & --- \\
    \bottomrule
    \multicolumn{10}{@{}l}{$^\dagger$Exploratory (\cref{sec:predictors}).}
  \end{tabular}
\end{table}

\begin{figure}[t]
  \centering
  \includegraphics[width=\linewidth]{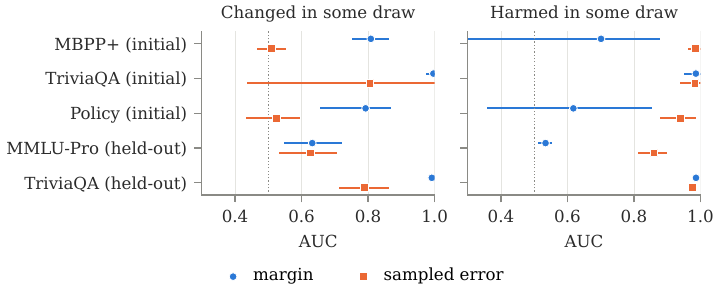}
  \caption{AUC of the two predictors at bit~6, with 95\% intervals. Only two to four items were harmed on the initial
    tasks, and three changed on initial TriviaQA.}
  \label{fig:auc}
\end{figure}

By AUC, the margin predicts changes and the sampled error predicts harm
(\cref{fig:auc}). For changed items, the margin predicts better than
the sampled error on every task but MMLU-Pro, where nearly every item
changes in some draw. For harmed items, the sampled error predicts
better than the margin on every task but TriviaQA, where most changed
items are also harmed.

\subsection{Form of Harmed Responses}\label{sec:form}

As in the conference version~\cite{yamamoto2026secrypt}, every bit-6
and bit-15 flip falls in the partial-divergence class, and bit~14
collapsed the responses at 2 of its 3 coordinates. Among harmed
responses, only the MMLU-Pro responses without a closing tag, 303 of the
659 harmful draws at bit~6, break the required format; on the other tasks
every harmed response kept the tag.

The harmed responses on both policy items begin with the required
refusal and then make a medical claim anyway: on one item they link a
stiff neck to the pillow or sleeping position, and on the other they say
that redness can indicate an infection needing treatment.

\subsection{Returning Requests}\label{sec:returning}

On returning requests, H2 also holds on both held-out tasks. Fresh
requests change more outputs than returning ones, but their harm rises
only on held-out TriviaQA (\cref{tab:fresh}): H3 holds on both held-out
tasks, and H4 only on held-out TriviaQA.

\begin{table}[t]
  \centering
  \caption{Returning (Ret.) and fresh requests on $D$ (\cref{sec:measures}),
    with the paired difference fresh minus returning and its 95\%
    interval. For policy, $h$ gives harmful draws, returning/fresh.}
  \label{tab:fresh}
  \scriptsize
  \setlength{\tabcolsep}{2.5pt}
  \begin{tabular}{@{}lrrrlrrl@{}}
    \toprule
    & & \multicolumn{3}{c}{$T$} & \multicolumn{3}{c}{$h$ ($\times 10^{-3}$)} \\
    \cmidrule(lr){3-5}\cmidrule(l){6-8}
    Task & Bit & Ret. & Fresh & Difference & Ret. & Fresh & Difference \\
    \midrule
    \multicolumn{8}{@{}l}{\textit{initial}} \\
    MBPP+    & 6  & 0.11   & 0.13   & 0.017 [$-$0.008, 0.043]     & 2.2 & 2.9 & 0.74 [$-$2.80, 4.87] \\
             & 15 & 0.11   & 0.13   & 0.028 [0.001, 0.055]        & 2.4 & 2.5 & 0.15 [$-$3.10, 3.69] \\
    TriviaQA & 6  & 0.0047 & 0.0094 & 0.0047 [0, 0.0138]          & 3.6 & 6.9 & 3.3 [$-$0.4, 11.6] \\
             & 15 & 0.0054 & 0.0065 & 0.0011 [$-$0.0043, 0.0076]  & 4.3 & 5.4 & 1.1 [$-$3.6, 7.2] \\
    Policy   & 6  & 0.18   & 0.17   & $-$0.0086 [$-$0.0603, 0.0453] & \multicolumn{3}{l}{12/12} \\
             & 15 & 0.19   & 0.17   & $-$0.021 [$-$0.073, 0.037]  & \multicolumn{3}{l}{8/9} \\
    \multicolumn{8}{@{}l}{\textit{held-out}} \\
    TriviaQA & 6  & 0.0026 & 0.0053 & 0.0028 [0.0013, 0.0045]     & 1.6 & 3.1 & 1.6 [0.5, 2.8] \\
             & 15 & 0.0038 & 0.0053 & 0.0015 [0.0001, 0.0030]     & 2.4 & 3.2 & 0.79 [$-$0.17, 1.84] \\
    MMLU-Pro & 6  & 0.60   & 0.64   & 0.041 [0.019, 0.065]        & 33  & 37  & 3.6 [$-$2.0, 9.3] \\
             & 15 & 0.62   & 0.65   & 0.025 [0.004, 0.045]        & 37  & 39  & 2.2 [$-$3.9, 7.9] \\
    \bottomrule
  \end{tabular}
\end{table}

%% file: 06_discussion.tex
\section{Discussion}\label{sec:discussion}

The measurements change how a silent flip should be counted
(\cref{sec:implications}), rated (\cref{sec:severity}) and countered
(\cref{sec:countermeasure}); \cref{sec:limitations} states their
limits.

\subsection{Implications for Measurement}\label{sec:implications}

The conference version concluded that the damage of a flip accumulates
with how long the block stays cached and that prefix blocks warrant
integrity protection~\cite{yamamoto2026secrypt}. Counted as harm, the
damage still accumulates, but at rate $h$, and the fraction of changes
that is harmful differs strongly by task: 0.023 on MBPP+, 0.060 on
MMLU-Pro and 0.59 on held-out TriviaQA at bit~6. Beyond staying
coherent~\cite{yamamoto2026secrypt}, harmed responses at bit~6 keep the
required format on every task except MMLU-Pro, where 356 of its 659
harmed responses do (\cref{sec:form}). They are thus even harder to catch by monitoring
outputs, which strengthens the case for protection
(\cref{sec:countermeasure}).

Over the $N$ requests that read a corrupted block, a flip harms
$h \times N$ responses, but change rates count all
$T \times N$ changed ones (\cref{tab:rates}). As a measure of harm,
they would also count changes that harm nothing, such as those from a
change in batch composition (\cref{sec:rq1}).

Moreover, nearly every harmful draw lands on a fragile item
(\cref{tab:fragile}). The harm of a flip in a deployment thus depends on
the share of fragile requests in its traffic. This share is not measurable from change rates but
measurable by sampling without injection, for requests whose responses
can be scored.

\subsection{Severity Assessment}\label{sec:severity}

Under the Common Vulnerability Scoring System
(CVSS)~\cite{first2019cvss}, we reassess three metrics of our advisory
(\cref{sec:ethics}):
\begin{description}
  \item[Attack vector: Physical to Local.] The co-located tenant induces
    the flip from its own code (\cref{sec:attacker}).
  \item[Availability: None to High.] A flip at a chosen bit can collapse
    the responses and so deny service to every request that shares
    the prefix (\cref{sec:surface,sec:scope,sec:form}).
  \item[Integrity: Low, unchanged.] The attacker controls neither which
    responses are harmed nor what they say, and a flip harms at most
    0.058 of them (\cref{fig:scatter}).
\end{description}
The vector becomes
\nolinkurl{CVSS:3.1/AV:L/AC:H/PR:L/UI:N/S:U/C:N/I:L/A:H}, and the base
score rises from 1.8 to 5.3. Flips at the silent bits alone leave
availability unaffected and would score 2.5.

\subsection{Countermeasure}\label{sec:countermeasure}

As \cref{sec:implications} shows, harmed responses look like ordinary
errors, and they fall on fragile items (\cref{sec:rq2}).
Integrity should therefore be checked at the cached block rather than
in the outputs. The countermeasure of the conference version computes a
checksum of each prefix block when the block is cached and verifies it
before each use; it detected all 120 injected flips without false
positives at negligible throughput cost~\cite{yamamoto2026secrypt}. It
limits the requests served from a corrupted block to one batch, except
for a flip that lands between a check and the use. The damage of one
flip thus falls from $h \times N$ to $h$ times the batch size.

\subsection{Limitations}\label{sec:limitations}

\paragraph{Internal validity: fragile items.} On MMLU-Pro, fresh
requests answer some fragile items incorrectly even without a flip, so
these items fall outside $C$. The fresh $h$ and the comparison of
request types (\cref{sec:returning}) thus cover fewer of the items on
which harm concentrates. Moreover, 16 samples miss rare errors and so
count some fragile items as non-fragile, which makes the test of H2
conservative; an item answered correctly with probability 0.95, for
example, is counted as non-fragile with probability 0.44.

\paragraph{Construct validity: scoring.} Two choices of scoring shape
$h$. First, alias matching counts misspelled or reworded answers,
such as ``Tamar River'' for ``Tamar'', as wrong, which overstates the
reported $h/T$ of held-out TriviaQA at bit~6: 0.59 (\cref{tab:rates})
instead of 0.47 by our reading. Second, on MMLU-Pro, $h$ includes
responses without the closing tag, which a format check would flag
(\cref{sec:form}); the harm that evades such a check is about half of
$h$.

\paragraph{External validity: setting.} Our measurements cover one
model, one prefix, single flips in the value cache at two bit
positions, and BF16. The conference version found change rates on
DeepSeek-R1-Distill-Llama-8B close to those on Qwen3-8B at bits~6
and~15~\cite{yamamoto2026secrypt}, but did not measure harm. The 100 policy requests were written
for this study, and all policy harm falls on two of them.

\paragraph{External validity: residency.} The damage $h \times N$ grows
with how long a corrupted block stays cached. The conference version
saw no eviction over 100 requests~\cite{yamamoto2026secrypt}. In one
large provider's API traces, 99\% of cached blocks were last reused
within 97 seconds, but blocks of system prompts, which the attack
targets, were reused for up to 1{,}200 minutes, so $N$ can be
large~\cite{wang2025kvcachewild}.

\paragraph{External validity: physical attacks.} Ideal bit targeting
(\cref{sec:scope}) relaxes the physical constraints of Rowhammer, such
as row adjacency, fixed flip direction and undocumented address
mappings~\cite{kim2014rowhammer,lin2025gpuhammer}. A physical attack
needs three steps:
\begin{itemize}
  \item \emph{Detecting that the target prefix is cached}, shown with
    side channels~\cite{song2025earlybirdleak,wu2025promptpeek,pennas2026cachesolidarity}.
  \item \emph{Flipping the chosen bit}, shown at bits~6 and~14 of
    16-bit values but not at bit~15~\cite{lin2025gpuhammer}.
  \item \emph{Placing the flip in a long-lived prefix block}, not
    shown: CacheTrap reached only the short-lived cache of a single
    request~\cite{alnahian2025cachetrap}.
\end{itemize}
The attack also needs the conditions of \cref{sec:attacker}. With ECC
enabled, as NVIDIA recommends~\cite{nvidia2025rowhammer}, most flips
become detectable errors: 94 against one silent corruption on an
A6000~\cite{lin2026gputhor,nvidia2026rowhammer}. No Rowhammer flips
have been shown on GPUs with HBM~\cite{lin2026gpubreach} or on the
GDDR6 data-center A10 and L4~\cite{lin2026gputhor}.

\subsection{Ethics and Disclosure}\label{sec:ethics}

All experiments injected flips in software into a locally hosted model,
without production systems, user traffic or hardware attacks. We
reported the vulnerability, with a proposed patch, to the vLLM
maintainers through a GitHub Security Advisory on 8 April 2026, before
the conference version was published~\cite{yamamoto2026secrypt}. The
maintainers declined it, considering the threat implausible in current
deployments.

The harness with its prompts and scoring rules, the item lists, the run
outputs and the hypothesis files will be released upon publication.

%% file: 07_related_work.tex
\section{Related Work}\label{sec:related}

Prior fault studies of LLMs measure impact as an aggregate error rate or
as a change rate. None asks what fraction of the changes is harmful or
which items the harm falls on; this paper answers both, with $h/T$ and
with predictors that need no injection.

\begin{table}[t]
  \centering
  \caption{What representative fault studies of LLMs measure. SDC: silent
    data corruption, a correct output turned incorrect; DUE: detected
    unrecoverable error.}
  \label{tab:related}
  \footnotesize
  \setlength{\tabcolsep}{3pt}
  \begin{tabular}{@{}lll@{}}
    \toprule
    Work & Fault target & Outcome measured \\
    \midrule
    Chen et al.~\cite{chen2025flipguard}  & weights, activations, gradients & accuracy, perplexity \\
    Chai et al.~\cite{chai2025gpusofterrors} & registers of GPU instructions & SDC and DUE rates \\
    Huang et al.~\cite{huang2026errorpropagation} & activations, weights, KV cache & benchmark accuracy \\
    AttentionBreaker~\cite{das2025genbfa} & weights & accuracy destroyed \\
    SBFA~\cite{guo2025sbfa} & weights & accuracy destroyed \\
    PrisonBreak~\cite{coalson2025prisonbreak} & weights & jailbreak success \\
    Conference version~\cite{yamamoto2026secrypt} & shared prefix block & output change \\
    \midrule
    This paper & shared prefix block & $h$, $h/T$; where harm falls \\
    \bottomrule
  \end{tabular}
\end{table}

\subsection{Bit-Flip Attacks on Neural Networks}\label{sec:rw-bfa}

Model weights are the best-studied target of bit-flip attacks.
Software studies under ideal bit targeting, starting with
BFA~\cite{rakin2019bfa}, search for the few weight bits whose flips
destroy a model's accuracy, and later work extended this search to
LLMs~\cite{das2025genbfa,guo2025sbfa}. Rowhammer has realized such
flips physically: to degrade DNNs in CPU DRAM~\cite{yao2020deephammer},
to implant a backdoor with a single flip~\cite{li2025oneflip}, and to
jailbreak aligned LLMs in GPU memory~\cite{coalson2025prisonbreak}.
Unlike a weight, which every inference reads, a shared prefix block
reaches only the requests that share its prefix.

\subsection{KV-Cache Security and Resilience}\label{sec:rw-kv}

The KV cache has been attacked mainly through side channels. Prefix sharing lets
a co-located tenant infer, from timing, whether a prompt is
cached~\cite{song2025earlybirdleak} and reconstruct other tenants'
prompts~\cite{wu2025promptpeek}. PrefixWall defends against the timing
channel~\cite{pennas2026cachesolidarity}, and the cache salt narrows
who shares a block (\cref{sec:surface}). Cache-key collisions can make a
request read another prompt's blocks~\cite{wu2026cacheme}, a software
route to an integrity violation that needs no fault.

Few studies alter the cached tensors. CacheTrap implants a trojan by
flipping a single value-side bit in the short-lived KV cache of one
request and validates the attack chain with a physical flip on an
A6000~\cite{alnahian2025cachetrap}. History
swapping assumes read and write access to the cache and replaces
contiguous spans of it with precomputed
ones~\cite{ganesh2025kvcachemanipulation}, a stronger capability
than one hardware fault. Huang et al.\ inject memory faults directly into the
KV cache during inference and find modest drops in benchmark accuracy,
but do not consider a block that many requests
share~\cite{huang2026errorpropagation}. Kelle studies non-adversarial retention errors in an
eDRAM KV cache for edge devices~\cite{xia2025kelle}. The conference
version of this paper introduced the shared prefix block as a fault
target and characterized how most of its flips change outputs without
collapsing them, reach only the requests that share it, and
persist~\cite{yamamoto2026secrypt}.

\subsection{Measuring the Impact of Faults}\label{sec:rw-eval}

\Cref{tab:related} lists what representative studies measure. The rate
of silent data corruption (SDC) comes closest to $h$: Chai et al.\
count, on tasks scored by accuracy, among inferences that are correct
without faults, those whose
output becomes incorrect~\cite{chai2025gpusofterrors}. None of these
measures conditions on change or shows which items the harm falls on.

%% file: 08_conclusion.tex
\section{Conclusion}\label{sec:conclusion}

A bit-6 or bit-15 flip in a shared prefix block changes many responses,
but the fraction of the changes that is harmful depends strongly on the
task: 0.023 on MBPP+, 0.060 on MMLU-Pro and 0.59 on held-out TriviaQA at
bit~6.
The damage that the conference version found to accumulate with block
residency~\cite{yamamoto2026secrypt} remains; what accumulates is
harm, at rate $h$ rather than $T$.
The harm falls on fragile requests, whose sampled answers are sometimes
wrong and which sampling without injection can find.

Fault studies of LLM serving should therefore report harm by task,
conditioned on change. In a deployment, the harm of a flip depends on
the share of fragile requests in its traffic, which can be measured for
requests whose responses can be scored. Because harmed responses mostly
keep the required format and are hard to tell from ordinary errors,
the case for the integrity protection that the conference version
called for grows stronger: a flip should be detected by checking
cached blocks rather than by monitoring outputs.

Two questions remain open (\cref{sec:implications,sec:limitations}):
how long a popular prefix block stays cached under production traffic,
which sets $N$, and how large the share of fragile requests is in real
workloads.